# Emergence of Chern insulating states in non-Magic angle twisted bilayer graphene

Cheng Shen（沈成）[1,2], Jianghua Ying（应江华）[1,2], Le Liu（刘乐）[1,2], Jianpeng Liu（刘健鹏）[3,4], Na Li（李娜）[1,2,6], Shuopei Wang（王硕培）[1,2,6], Jian Tang（汤建）[1,2], Yanchong Zhao（赵岩翀）[1,2], Yanbang Chu（褚衍邦）[1,2], Kenji Watanabe[7], Takashi Taniguchi[8], Rong Yang（杨蓉）[1,5,6], Dongxia Shi（时东霞）[1,2,5], Fanming Qu（屈凡明）[1,2,6], Li Lu（吕力）[1,2,6], Wei Yang（杨威）[1,2,6*] and Guangyu Zhang（张广宇）[1,2,5,6*]

[1]Beijing National Laboratory for Condensed Matter Physics; Key Laboratory for Nanoscale Physics and Devices, Institute of Physics, Chinese Academy of Sciences, Beijing 100190, China

[2]School of Physical Sciences, University of Chinese Academy of Sciences, Beijing 100190, China

[3]School of Physical Sciences and Technology, ShanghaiTech University, Shanghai 200031, China

[4]ShanghaiTech laboratory for topological physics, ShanghaiTech University, Shanghai 200031, China

[5]Beijing Key Laboratory for Nanomaterials and Nanodevices, Beijing 100190, China

[6]Songshan-Lake Materials Laboratory, Dongguan, Guangdong523808, China

[7] Research Center for Functional Materials, National Institute for Materials Science, 1-1 Namiki, Tsukuba 305-0044, Japan

[8]International Center for Materials Nanoarchitectonics, National Institute for Materials Science, 1-1 Namiki, Tsukuba 305-0044, Japan

*Corresponding author: wei.yang@iphy.ac.cn; gyzhang@iphy.ac.cn

**Abstract:** Twisting two layers into a magic angle (MA) of ~1.1º is found essential to create low energy flat bands and the resulting correlated insulating, superconducting, and magnetic phases in twisted bilayer graphene (TBG). While most of previous works focus on revealing these emergent states in MA-TBG, a study of the twist angle dependence, which helps to map an evolution of these phases, is yet less explored. Here, we report a magneto-transport study on one non-magic angle TBG device, whose twist angle $\theta$ changes from 1.25º at one end to 1.43 º at the other. For $\theta = 1.25°$, we observe an emergence of topological insulating states at hole side with a sequence of Chern number $|C| = 4 - |v|$, where $v$ is the number of electrons (holes) in moiré unite cell. When $\theta > 1.25°$, the Chern insulator from flat band disappears and evolves into fractal Hofstadter butterfly quantum Hall insulator where magnetic flux in one moiré unite cell matters. Our observations will stimulate further theoretical and experimental investigations on the relationship between electron interactions and non-trivial band topology.

**Main text**

Twist angle plays an important role in TBG and other two-dimensional twisted moiré systems. By twisting two layers into a magic angle (MA) of ~1.1º in TBG, moiré bands are generated with a narrow band width of ~10 meV, so called flat band where it favors electron interactions over kinetic energy[1-9], contributing to the realizations of correlated insulating[2], superconducting[3,4,5], and orbital magnetic phases[5,10-12]. As the twist angle is increased further to non-magic angle (NMA), the moiré band width is increased very fast and the bands are no longer flat, where the electrons tend to lose correlation[1,13,14]. Recently, aside from the observation of non-zero Chern bands in TBG[5,11,12], twisted multilayer graphene[15,16] and ABC-stacked trilayer graphene[17] where inversion symmetry ($C_2$) is broken, the strong electron interactions in MA-TBG are found able to break time reversal symmetry (*T*), untie the degeneracy between the flat bands, and thus reveal a sequence of non-trivial topological insulating states with Chern number $|C| = 4 - |v|$[18-21], where $v=n/n_0$ is the number of electrons (or holes) filled in one moiré unit cell, *n* is the carrier density, and $n_0$ is carrier density of one electron per moiré unit cell area. Then, intuitively one may ask how strong an electron interaction is needed to produce these Chern bands, and if it survives even for NMA-TBG or not. The question is interesting and important, and yet to be explored.

In this paper, we try to bridge the gap by studying magneto transport properties of a NMA-TBG device with different twist angles. A perpendicular magnetic field (*B*) would have impacts on TBG in two different ways. First, the magnetic field tends to recombine the flat bands of TBG into a series of fractal Landau levels (LLs), i.e., the so called Hofstadter butterfly spectra, in which the recurring fractal bands are dependent on the number of magnetic fluxes in each moiré primitive cell. Second, aside from the formation of LLs, the magnetic field also induces splitting between the flat bands with opposite Chern numbers +/-1 due to the orbital Zeeman effect[22]. Such orbital Zeeman splitting originates from the intrinsic orbital magnetism of the topological flat bands of TBG, and can be dramatically enhanced by Coulomb interactions, which would give rise to a series of time-reversal broken moiré Chern bands without the necessity of forming *LLs*. (*Note that quantum Hall insulator is in principle a Chern insulator, where the LL filling factor corresponds to the Chern number. To avoid confusion, we use "Chern" specifically for moiré Chern bands in this paper.*) These moiré Chern bands under (weak) perpendicular magnetic field *B* are manifested by quantized Hall conductance $\sigma_{xy} = Ce^2/h$, where *e* is the electron charge and *h* is the Planck's constant. The Chern number could also be obtained from a linear slop of the quantized conductance in landau fan diagram by $C = (h/e)(dn/dB)$, where *n* is the carrier density. Although the ratio between the characteristic Coulomb interaction strength *U* and the moiré bandwidth *W* is significantly reduced for NMA TBG, it is still of fundamental interest to reveal the interplay and competition between the fractal LL spectra and the topological moiré bands driven by (interaction-enhanced) orbital Zeeman splittings. We shed light on such an intriguing problem by measuring Landau-fan diagrams in NMA TBG samples with different twist angles.

A schematic illustration of the device is shown in Figure 1a. The twisted graphene layers are encapsulated by hBN, and they are not aligned with hBN substrate in order to hold the $C_2$ sublattice symmetry (Fig. S1). The non-Magic angle nature is revealed in the transfer curves at different positions as shown in Fig. 1b, from which the extracted twist angle changes from 1.25° at one end to 1.43° at the other in Fig.1c, and correlated insulating states are absent in our device. The twist angles are further quantified and validated in the magneto-transport, regardless of a moderate inhomogeneity of ±0.03° for each device presented in the main text. The details for assignment of twist angle and related discussions of twist inhomogeneity are presented in the first section of Supplementary.

We start with magneto transport studies at a twist angle $\theta$ = ~1.38°, where fractal Hofstadter butterfly spectra dominate. The measurements are performed at 20mK, and a color mapping of Hall resistance $R_{xy}$ as a function of carrier density and magnetic field is shown in Fig. 1d. In TBG moiré system, a single particle picture of fractal LL spectra can be derived from the interplay of periodic interlayer moiré potential and magnetic fields[23]. They are featured by the recurring embryo LLs which emanate from magnetic flux per moiré unit cell $\Phi = \Phi_0(p/q)$, where *p* and *q* are co-prime integers, $\Phi_0 = h/e$ is the quantum magnetic flux. These embryo LLs are well understood as replica minibands at effective magnetic field $B_{eff} = \pm|B - B_{p/q}|$, with spin-valley symmetry breaking driven by electron interactions[24-28]. Notably, as depicted in Fig. 1e, one could identify LLs from charge neutral point (CNP) at $v = 0$ and that from full filling at $v = 4$, and the fractal LLs appear only in the regimes between two rational fluxes at high magnetic field where LLs from $v=0$ and that from $v=4$ intersects, consistent with the Hofstadter butterfly spectra in the previous well-studied graphene/hBN superlattice heterostructure[24-28]. Additionally, quantum Hall ferromagnets (QHFM), well developed here with LL filling factors $v_{LL}=\pm1, \pm2, \pm3, 5, 6, 7$ near charge neutral point (CNP), are formed by lifting the 8-fold spin, valley, and layer degeneracies. Close to the Hofstadter gaps at $\Phi = \Phi_0/q$, the system experiences a reverse Stoner transition due to superlattice-modulated bandwidth broadening, which is manifested by the suppression of QHFM. Instead of being quantized plateaus, the Hall resistances develops into a series of peaks and valleys with positive and negative values respectively, and the alternating Hall resistance sign changes in the red curve of Fig. 1f is a direct evidence of recurring fractal Hofstadter Butterfly bands.

Next, we discuss the transport measurements at $\theta = \sim 1.25°$, an angle closer to magic angle of ~1.1°, and we observed a phases diagram beyond Hofstadter butterfly spectra. Similar to the device with $\theta = \sim 1.38°$, correlated insulating states are absent at zero magnetic field, and LLs from $v = 0$ and that from $v = 4$ at 20mK are clearly shown in Fig. 2a. What make the major difference is the emergence of longitudinal resistance $R_{xx}$ minima, depicted as the red dashed lines in Fig. 2b. These states are characterized by a series of Chern number (or LL filling factor) = -1, -2, -3, which could be traced to integer moiré band fillings of $v$ = -3, -2, -1 at $B = 0$ T, respectively. For simplicity, we name these states by ($C, v$), that is (-1, -3), (-2, -2), and (-3, -1).

These states are not originated from the fractal bands in Hofstadter Butterfly spectra. Take the state (-2, -2) for example, the onset magnetic field is very low, at $B < 2T$ ($\sim 0.05\Phi_0$), which suggests magnetic flux plays a trivial role (Fig.2a and Fig.S6). Besides, the state threads through several rational fluxes in a fan diagram, which is also in contrast to the expected restriction to one rational flux for fractal minibands. Similar threading behaviors are found in the states (-1, -3) and (-3, -1), which emerge at $B$=8T and 12T respectively. It is also noted that the state (-2, -2) is robust, and it shows a thermal activation gap comparable to LL ($v_{LL}$ =-2) originated from CNP (Fig.S4). The twist angle inhomogeneity here precludes well-developed Hall plateaus for these states (Fig.S5).

We interpret these states as the emergence of topologically nontrivial moiré Chern insulators, which have also been observed in MA-TBG in recent related works[18-21,29,30]. As discussed in previous theoretical works, the low-energy flat bands in TBG can be interpreted as 8 valley-spin degenerate Chern bands with opposite Chern numbers +/-1, which carry opposite orbital magnetization and exhibit opposite orbital $g$ factor[22]. As a result, once a (weak) perpendicular magnetic field is applied, time-reversal symmetry would be broken due to the orbital Zeeman effect, which splits the 8-fold degenerate flat bands into two sets of 4-fold degenerate Chern bands with $C$=+1 and $C$=-1 respectively, equivalent to zeroth pseudo LLs[31]. If an integer number ($v$) of the 4-fold degenerate Chern bands are filled (emptified) on the electron (hole) side, a gap would be opened up between the occupied and unoccupied bands due to exchange interactions, leading to an interaction-driven Chern insulator with Chern number $|C| = 4 - |v|$. Therefore, one would expect to see a sequence of Chern insulator states (-1,-3), (-2,-2), (-3,-1) as the filling factor $v$ decreases from -1 to -3, which is clearly marked in Fig.2b. Such a picture is similar to the QHFM for cyclotron LLs in graphene[32,33]. In a finite magnetic field, Coulomb interactions tend to first break either one of the spin or valley symmetry forming twofold degenerate LLs, and eventually lift all fourfold degeneracy. Combing back to the flat bands of TBG, the above picture implies that a symmetry-broken Chern insulator (SCI) (-2, -2) should appear first with the increase of magnetic field, followed by the (-1,-3) or (-3, -1) SCI states. This argument is consistent with our observed hierarchy of SCIs, and explains why a tiny field (~2T) is required to establish the (-2,-2) state. It is worthwhile to note that although an approximate particle-hole symmetry is present in the continuum model of TBG, it is absent in a more realistic band structure of TBG including the effects of atomic corrugations[34], in which the conduction flat band has a wider bandwidth than the valence band, which suppresses the emergence of the SCIs on the electron side.

Our results demonstrate the emergence of non-trivial band topology is not restricted to the magic angle regime. Aside from the $R_{xy}$ data acquired at T=20mK for $\theta$=1.38°, we also measured $R_{xx}$ responses for various $\theta$ of 1.25°, 1.38° and 1.43° at an elevated temperature $T$=1.7K, with B=0-9T. For band filling -4< $\nu$<0, SCI (-2, -2) and (-1, -3) are observed in $\theta$=1.25° TBG device, but not the $\theta$=1.38° and 1.43° devices (see Fig. 1, Fig. 3, Fig. S7, and Fig. S8). Experimentally, it is surprising to find the SCI in 1.25° device for two reasons, one is the non-magic twist angle, and the other is a moderate twist angle inhomogeneity. Usually, the inhomogeneity would destroy the correlated behavior and impede the observation of SCI. However, seeing is believing, the observation of SCIs in turn suggest the relatively big twist angle does not play a dominant role here, and an explanation of microscopic mechanism behind as well as what role played by strain is beyond our work. And instead, we tend to offer a tentative yet quantitative analysis of electron interaction and kinetic energy, we found that our observation of Chern bands at 1.25 degree is very close to the crossover regime where Coulomb interaction and moiré band bandwidth are comparable (See details in Supplementary Fig. S9). Since the moiré bandwidth $W$ increases almost linearly with the twist angle (when $\theta$ is larger than the MA), the disappearance of the SCI states at larger twist angles can be explained by the reduced interaction effects due to the enhanced kinetic energy. Note that the above analysis electron interaction is kind of rough, and it strongly depends on the dielectric constant and also on how the electron separation is treated. The critical twist angle below which such SCIs can emerge requires more detailed exploration in the future.

As discussed above, the SCI states observed in the 1.25° device is interpreted as an interaction-driven symmetry breaking state triggered by a tiny onset magnetic field. The gap in the SCI state is generated by electrons' Coulomb interactions, while the topological nature of the gap is triggered by a tiny $B$ field through the orbital Zeeman coupling. Thus, the resulted symmetry breaking state exhibits an orbital ferromagnetic order with nonzero Chern numbers. At larger twist angles, the Coulomb interaction may not be strong enough to overcome the moiré kinetic energy, then the symmetry-breaking scenario sketched above cannot happen and the system would still valley and spin symmetries (despite the non-interacting spin and orbital Zeeman splittings induced by the small $B$ fields, which are negligible compared to the bandwidth without considering interaction effects). However, another possible scenario is that the absence of SCI states in 1.38° and 1.43° TBG is accounted by the intrinsic trivial topology of the low-energy bands at larger twist angles such that the system remains topologically trivial even if the interactions are strong enough to drive the system into a symmetry-breaking state at larger twist angles. Although the intervalley couplings at the single-particle are still negligible (~0.05-0.1meV) at 1.38° and 1.43°, it could be dramatically enhanced by electron-phonon coupling[35], thus we cannot rule out this possibility. A method to distinguish these two scenarios is further elevating magnetic fields to enhance exchange Coulomb energy which is proportional to $\sqrt{B}$ or reducing device disorders, to fully break spin and valley isospin symmetry. If the nontrivial topology is preserved at larger twist angles, one would expect to see the emergence of SCIs with fully lifted valley-spin degeneracy at stronger magnetic fields. The specific mechanism for the interplay between electrons' interactions and the nontrivial band topology would be further elucidated by checking the SCI states would emerge or not at high-enough magnetic fields.

Lastly, we show the interplay and competition between these Chern bands, cyclotron LLs and also Hofstadter minibands in magnetic fields. In fact, fractal Hofstadter Butterfly spectra are greatly suppressed with $\theta$ = ~1.25° in Fig. 2a. Usually a relatively smaller twist angle gives a longer moiré wavelength and a stronger moiré potential, which should help to develop the fractal bands. However, instead the color mapping in Fig. 2a yields a Landau fan diagram with almost no trace of fractal bands fanning out at a rational filling of quantum magnetic flux, which is beyond the framework of single particle Hofstadter Butterfly picture as demonstrated in Fig. 1d and 1e with $\theta$ = ~1.38°. This argument is further supported in Fig. 3d and 3e at elevated temperature of T = 1.7K. The fractal Hofstadter butterfly survives as the Brown-Zak oscillations[36] in Fig. 3e for $\theta$ = ~1.38°, while they are absent in Fig. 3d for $\theta$ = ~1.25°. The suppression of conventional fractal Hofstadter bands suggests a competition, which is sensitive to electron interactions, between LL quantization effect and non-trivial topological effect at zero magnetic field from moiré Chern bands.

Here, we give a simple and self-consistent qualitative explanation in the framework of a competition between orbital Zeeman effect and cyclotron quantization. In one way, the orbital Zeeman effects from the nontrivial topology of the flat bands in TBG are characterized by effective orbital g-factor[37-39], which might be enhanced by electron-electron interactions. In another way, the cyclotron quantization is proportional to fermi velocity (or inversely proportional to effective mass). At a magic angle, the moiré band is ultra-flat with the fermi velocity greatly suppressed, i.e. orbital effects dominate over cyclotron quantization, and thus moiré Chern bands are dominating in the Landau fan diagram with a greatly suppressed LL quantization and eventually suppressed fractal Hofstadter butterfly spectra. When the twist angle increases to a non-magic angle, the fermi velocity tends to increase and thus gives a situation where both orbital effects and cyclotron quantization matters. As a result, moiré Chern bands and cyclotron LLs coexist and show intermediate interactions, as shown by instances that the trajectory of SCI (-2, -2) beads between rational fluxes $p\Phi_0/10$ and is crossed by Landau fan emanating from CNP (Fig.2b). At some point, e.g. by increasing the twist angle or by tuning doping level, the orbital effects fail and cyclotron quantization will win, eventually moiré Chern bands give way to Hofstadter butterfly fractal gaps, as manifested by prevailed Brown-Zak oscillations in electron branch of 1.25° NMA-TBG and also both electron and hole branches of 1.38° NMA-TBG. For a 1.43° device, the disappearance of moiré Chern bands is strongly related to the reduced *U/W* for an increased moiré bandwidth, and the disappearance of Brown-Zak oscillations is due to a smaller moiré period and weaker moiré potential, in line with fan diagram of a 1.8° TBG device in ref.13 where two graphene layers are weakly coupled.

In summary, our results demonstrate non-trivial topology for low-energy flat bands in non-magic-angle TBG with broken *T* symmetry. The stabilization of Chern insulators requires both a tiny magnetic field and strong electron interactions stemmed from flat bands, implying a related *T* symmetry breaking mechanism. Our studies also point out a crucial role of electron interactions in shaping Landau level phases in TBG. These discoveries would help us to unveil the mystery of electron correlation effects in band topology and even correlated insulators and superconductivities.

**Acknowledgements**

We appreciate the helpful discussion with Q. Wu and Y. Guan. The authors thank the finical supports from the National Key R&D program (Nos. 2020YFA0309604), NSFC (Nos. 61888102, 11834017, and 12074413), the Strategic Priority Research Program of CAS (Nos. XDB30000000 and XDB33000000), the Key-Area Research and Development Program of Guangdong Province (Grant No.2020B0101340001), and the Research Program of Beijing Academy of Quantum Information Sciences under Grant No. Y18G11. J. L. acknowledges the start-up grant of ShanghaiTech University and the National Key R&D program (No. 2020YFA0309601). K.W. and T.T. acknowledge supports from the Elemental Strategy Initiative conducted by the MEXT, Japan, Grant Number JPMXP0112101001, JSPS KAKENHI Grant Number JP20H00354 and the CREST(JPMJCR15F3), JST.

**Author Contributions**

C.S., W.Y. and G.Z. conceived the project. C.S. fabricated the devices and performed the transport measurements above 1.5K. C.S., J.Y., F.Q. and L.L. performed the transport measurements in dilution refrigerator. L.L. provided continuum model calculations. K.W. and T.T. provided hexagonal boron nitride crystals. C.S., W.Y. and G.Z. analyzed the data. C.S., J.L., W.Y. and G.Z. wrote the paper. All authors discussed and commented on this work.

**Figures and Figure Captions**

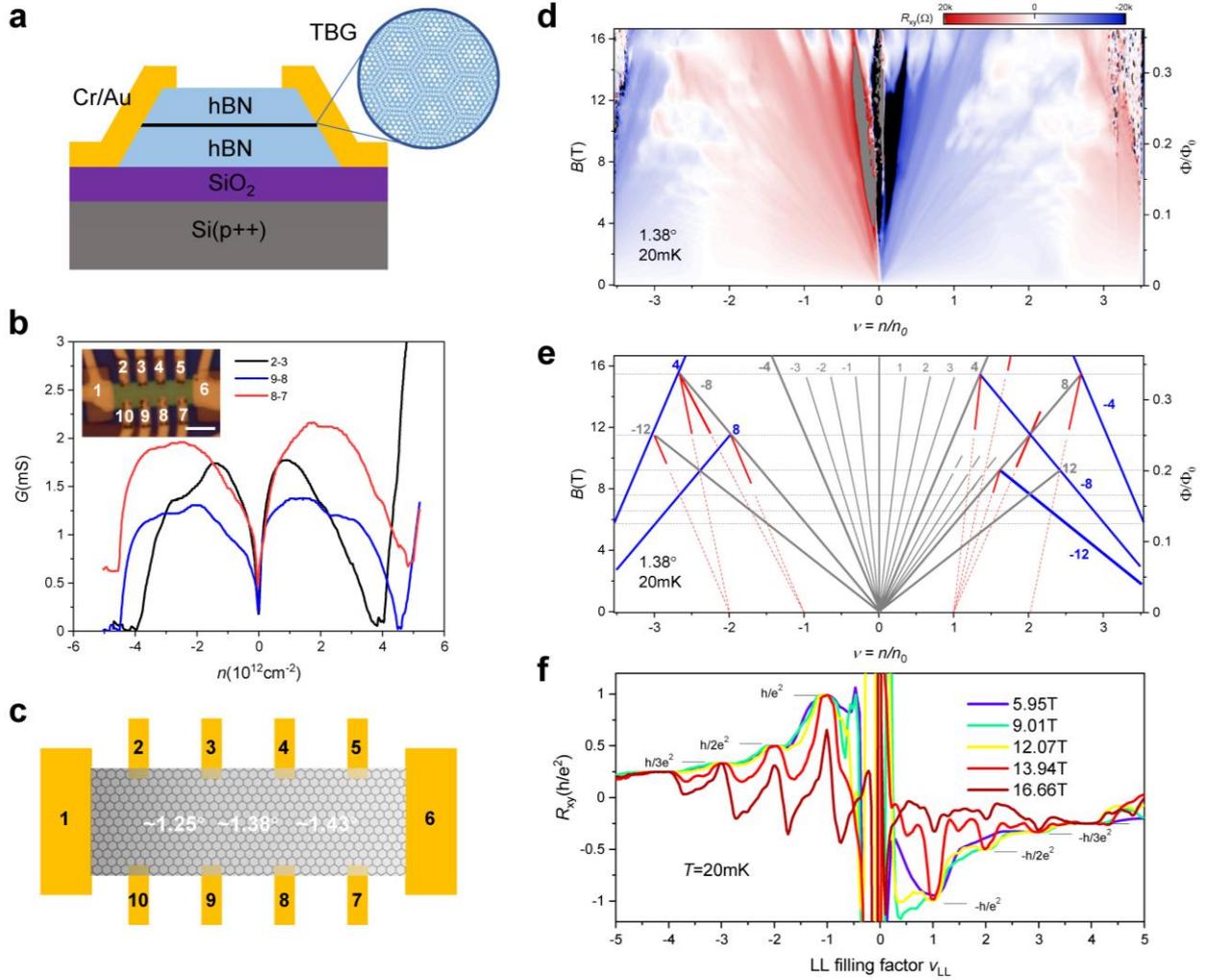

**FIG.1** Hofstadter butterfly spectra in TBG with $\theta=1.38°$. **a**, schematic side view of TBG device structure. **b**, two-terminal conductance as a function of carrier density $n$ acquired among different Hall bar pairs and at temperature $T=1.7$K. The scale bar in the inset is 4 μm. **c**, schematics of twist angle distribution. **d**, transverse Hall resistance $R_{xy}$ mapping plot versus electron or hole filling of moiré unit cell and perpendicular magnetic fields. **e**, Schematic illustration of Wannier diagram in **d**. **f**, line cuts at varied fields from **d**, showing suppression of quantum Hall ferromagnetism near CNP.

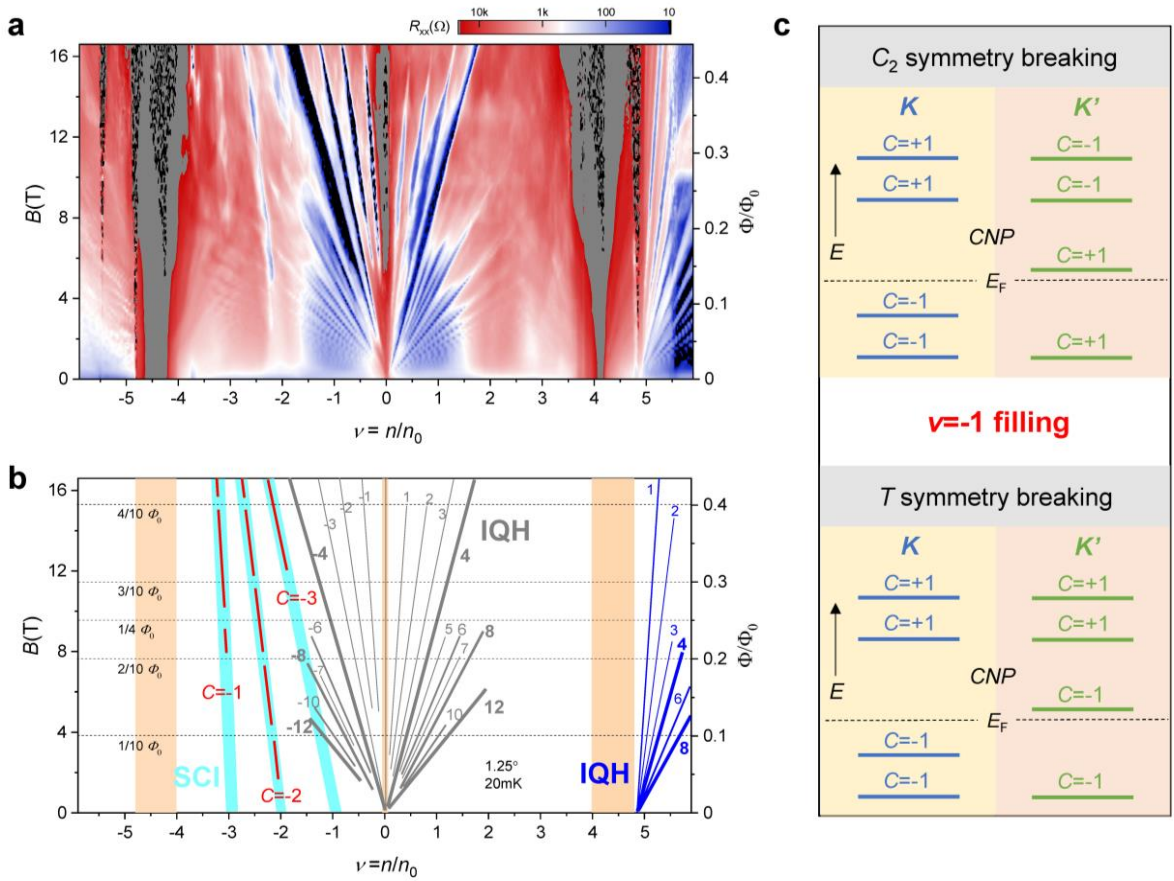

**FIG.2** Symmetry-broken Chern insulators in TBG with $\theta=1.25°$. **a**, Landau fan diagram of TBG acquired at temperature $T$=20mK. **b**, Schematic illustration of Wannier diagram. Grey and dark blue lines represent integer quantum Hall (IQH) insulators originated from CNP and gap edge of fulfilling, respectively. Light blue line shows symmetry-broken Chern insulators (SCI) interrupted by fractal Hofstadter gaps at rational magnetic flux. Orange shades represent band gap at CNP and fulfillings. **c**, Schematic of Chern number texture of flat band under $C_2$ symmetry or time-reversal symmetry breaking. The dash lines indicate Fermi level location at $v$=-1 filling.

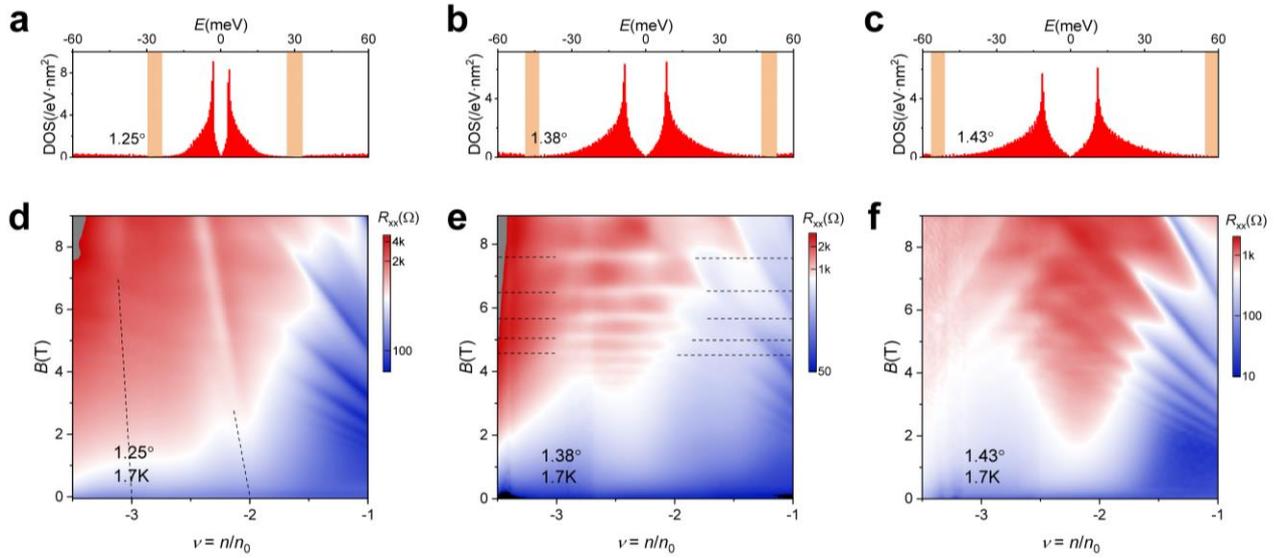

**FIG.3** Landau fan diagram in TBG with respect to twist angle $\theta$. **a**, **b**, **c**, Density of states (DOS) versus band energy for varied twist angles $\theta$=1.25°, 1.38° and 1.43°, respectively, according to Bistritzer-MacDonald model[1] calculations with a hopping energy of 110.7meV. Orange shades represent single-particle band gap at $v=\pm 4$ fillings. **d**, **e**, **f**, Landau fan diagrams for varied twist angles in one device. Dash lines in **a** and **b** show $R_{xx}$ minima of SCI and Brown-Zak oscillations, respectively.

# Supplementary Information

## 1. Device fabrication, twist angle distribution and band gap at CNP

The hBN/TBG/hBN sandwich structure is fabricated through a typical "tear and stack" method[40] and then followed by ebeam lithography and reactive ion etching techniques. One-dimensional Cr/Au edge contact is applied[41]. We misalign graphene and hBN substrate to preserve $C_2$ symmetry, and also reserve untwisted monolayer graphene where we can check whether the $C_2$ symmetry is truly kept according to its transport behavior. An absence of satellite resistance peaks in the measured density range for monolayer graphene reveals little moiré superlattice potential modulation present between graphene and hBN (Fig. S1b).

We extract roughly the dominant twist angle between each Hall bar pairs by the formula $n_s = \frac{8\theta^2}{\sqrt{3}a^2}$, which relates the twist angle $\theta$ to carrier density $n_s$ of full filling (here a is the graphene lattice constant). A further analysis of Hofstadter butterfly features, for instance Brown-Zak oscillations and fractal minibands, which directly give moiré unit cell area $A$ according to $\phi=BA=\phi_0/q$ (here $\phi$ is the featured rational magnetic flux, $q$ is an integer, and $B$ is the corresponding magnetic field), finally help us to define the exact value of twist angles. For 1.25 degree device, we extracted the twist angle from the carrier densities at which C=-3, -2, -1 Chern insulators at B=0T. And for 1.38 degree device, the twist angle is extracted from the Brown-Zak oscillations of Hofstadter butterfly due to the rational filling of magnetic flux in a moiré unit cell to quantum flux. For 1.43 degree device, the twist angle is quantified from carrier density difference between two points at B=0T where two set of landau levels fan out.

Fig. S1c shows temperature-varied resistance behavior in TBG with θ=1.25°. A metal-insulator transition occurs at low temperature, indicating gap opening at CNP. The linearly fitted thermal activation gap is about 1.35meV. Gap opening at CNP is typically attributed to $C_2T$ symmetry breaking[42], which is specifically induced by a $C_2$ symmetry or $T$ symmetry breaking. As $C_2$ symmetry is most likely already kept in our device, we thus speculate incipient $T$ symmetry breaking present in our device.

We summarize the transfer curves for all pairs of probes on both sides of the hall bar in Fig. S2. For probe pairs 2-3 and 10-9, which are located at different sides but a same distance from probe 1, they show similar carrier density for fulfilling, i.e. the similar twist angle, yet with different twist inhomogeneity (±0.03° for 2-3 and ±0.08° for 10-9). Results are the same for probes 3-4 and 9-8, as well as 4-5 and 8-7, with a moderate twist angle of ~±0.03°. The twist angle is increased when probes are further away from probe 1. In fact, such a twist angle distribution is most likely caused by the presence of bubbles at probes 1 and 6, which produce twist angle gradient from 1 to 6. Our transport data thus show an average result over a specific twist angle range. Despite the moderate twist angle inhomogeneity, it is noted that we can extract the exact twist angle contributed to the transport data precisely from the magnetic flux $\phi_0/q$ for Brown-Zak oscillation and carrier density at which Chern insulators are traced to at B=0T.

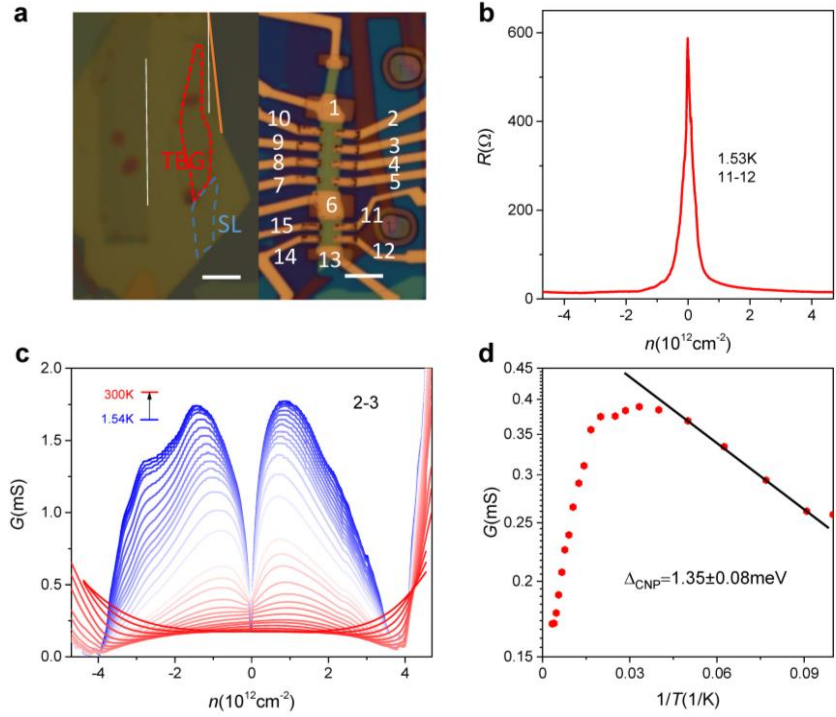

**FIG.S1** Symmetry breaking in TBG device. a, optical microscope images of our device. The left one shows initial sandwich image. We outline the edge direction of hBN and graphene with orange and white lines, respectively. Red and blue dash lines enclose regions of TBG and monolayer graphene, respectively. Scale bars of both images represent 4μm. b, four-terminal resistance as a function of carrier density in monolayer graphene. c, temperature dependence of four-terminal conductance acquired between electrode 2 and 3. d, Arrhenius fitting of thermal activation gap at CNP.

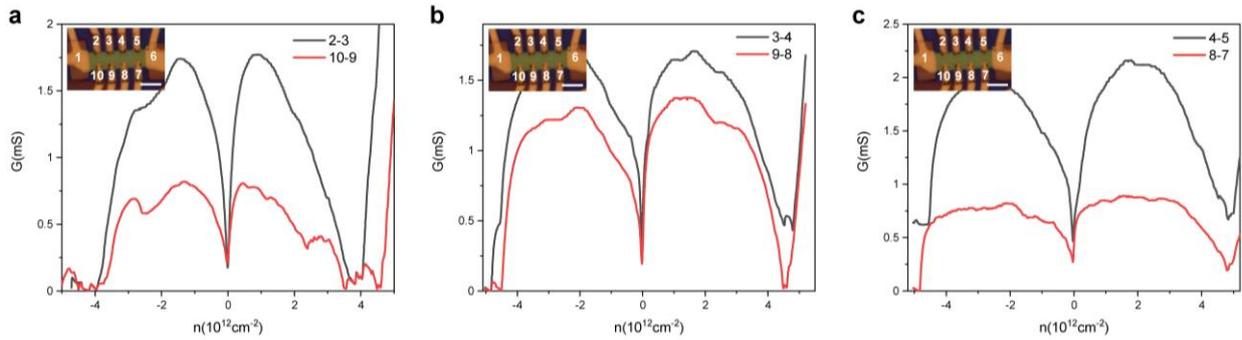

**FIG.S2** Twist angle distribution all over the device. From the carrier density where gap edge appears for fulfilling, we extract the twist angle these probes as ~1.25°(a), ~1.38° (b) and 1.43° (c). The data are acquired at 1.7K.

Generally speaking, mixed twist angle or twist angle inhomogeneity will tend to broaden the resistance peak at full filling, or even yield several peaks and thus bring several set of landau levels in the fan diagram (noted that the presence of gap opening at full filling will also tend to increase the width of the resistance peak). In our transport data, we only see two set of landau level, one from the charge neutral point, and the other from moiré superlattice period, which suggests trivial role played by twist angle inhomogeneity and a good sample quality of our device.

Besides, mixed twisted angle may bring additional scattering in the TBG and make it difficult to observe well quantized hall conductance plateau. However, the failure to observe quantized conductance in our 1.25 degree device is more likely due to big bandwidth at non-optimum twist angle, rather than few meV at magic angle of ~1.1 degree.

Lastly, mixed twisted angle is also predicted to induce $C_3$ symmetry breaking and lift landau level degeneracy in TBG devices[43]. This might relate to the lifted degeneracy transition from 8 folds to 4 folds for Landau levels emanating from charge neutral point, indicated by the dashed line in the figure below. However, electron interaction might also lift the degeneracy of landau levels.

## 2. Emergence of symmetry-broken and fractional Landau levels in remote bands

While most of attention is focused on the flat band in TBG, characteristics of remote dispersive bands are rarely studied. For remote bands, Fermi surface encloses Γ point in moiré Brillouin zone, producing four-fold spin-valley degeneracy. We find here that the four-fold degeneracy is fully lifted, yielding well-developed symmetry-broken LLs with $v_{LL}$=1, 2, 3 (Fig. S3). These LLs are characterized by longitudinal magnetoresistance $R_{xx}$ minima around zero, but without Hall plateaus due to twist angle inhomogeneity. We further find two $R_{xx}$ minima trajectories with a slope $\frac{dB}{dn} = \frac{\phi_0}{v_{LL}}$, where $v_{LL}$ for one LL is estimated to be 1/3, and another uncertainly to be 2/3 or 3/5. This phenomenon indicates an emergence of fractional quantum Hall effect (FQHE) for remote bands.

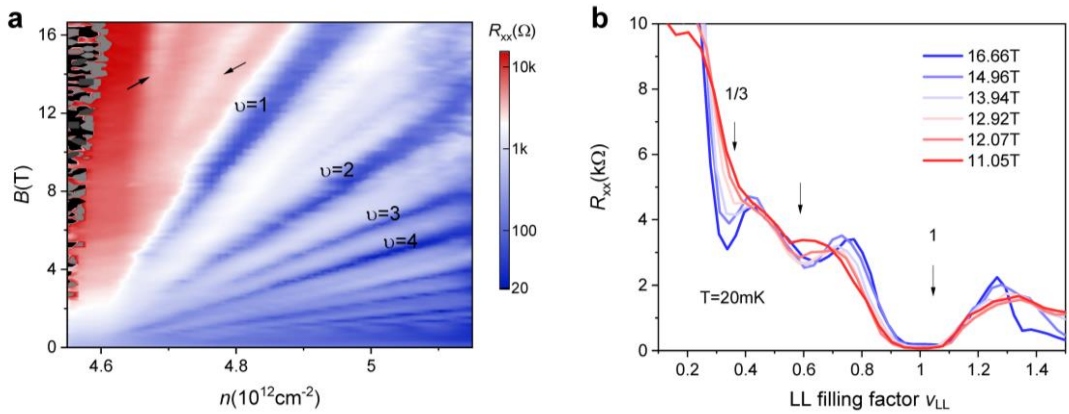

**FIG.S3** Symmetry-broken Landau levels and signatures of fractional quantum Hall effects in remote dispersive bands in $\theta$=1.25° TBG. The arrows in (a) point to $R_{xx}$ minima of fractional Landau levels, as depicted in (b).

## 3. Chern insulators: gap size, Hall resistance and onset at tiny field

Because of the pronounced twist angle inhomogeneity in our device, Hall resistances for Chern insulators (Fig. S5) deviate from quantized plateaus, instead shows a peak tracing the corresponding $R_{xx}$ minima as in Fig. 2a. We apply Arrhenius fitting to obtain the thermal activation gap for Chern insulator (-2, -2). At B=9T, its gap reaches to 11.7K, which is comparable to that of LL with $v_{LL}$=-2 from CNP. We also zoomed in the LL fan diagram to focus on the onset of Chern insulator (-2, -2). Fig. S6 shows that the $R_{xx}$ minima and also sign change of d$R_{xx}$/dn for Chern insulator (-2, -2) develop below B=1T.

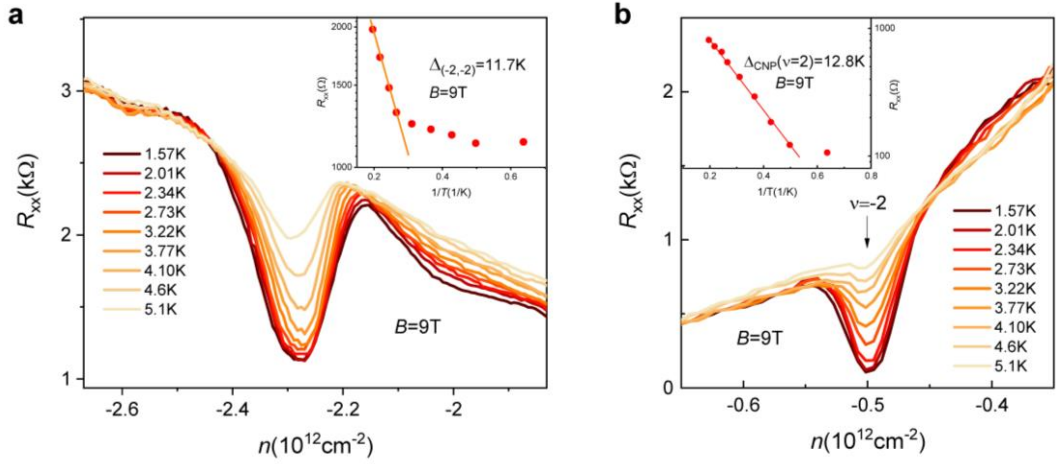

**FIG.S4** Temperature dependence of $R_{xx}$ at B=9T for Chern insulator (-2, -2) (a) and symmetry-broken LL $\nu_{LL}$=-2 originated from CNP (b). The inset figures show fitted thermal activation gap according to Arrhenius formula $R \propto \exp(-\Delta/2kT)$.

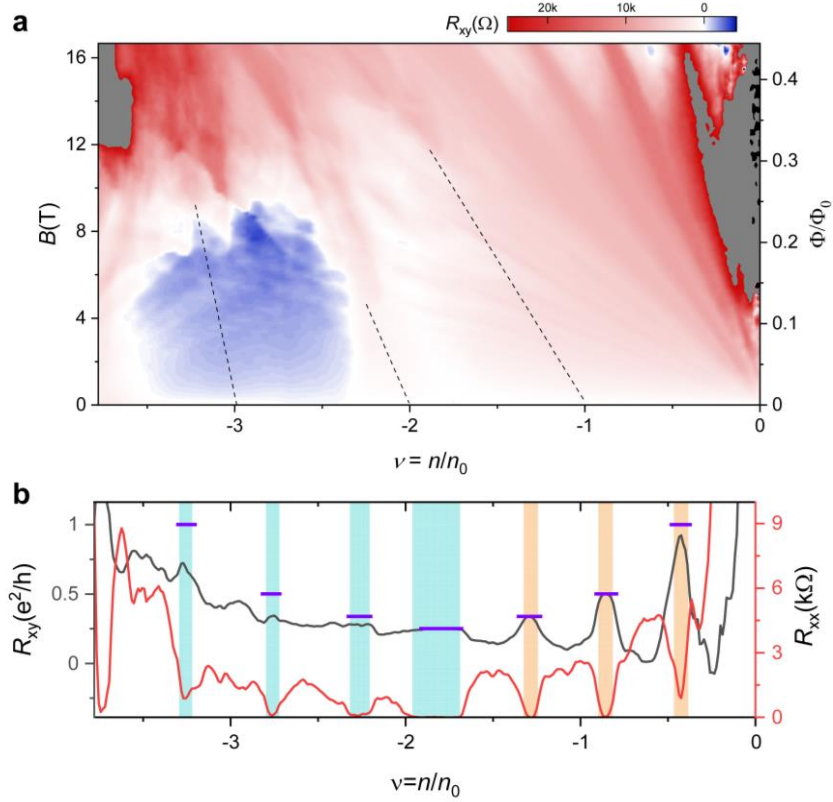

**FIG.S5** Hall resistance $R_{xy}$ behaviors in $\theta$=1.25° TBG. a, two-dimensional mapping of $R_{xy}$ as a function of carrier density and magnetic fields. Dash lines show expected trajectories of Hall plateaus and $R_{xx}$ minima for Chern insulators. b, line cuts of $R_{xy}$ and $R_{xx}$ at B=16.66T. We mark the Chern insulators (-1, -3), (-2, -2), (-3, -1) and (-4, 0) with light blue shades and symmetry-broken LLs from CNP with orange shades. Additionally, violet bars are added to show the expected quantized value of $R_{xy}$. All the data are acquired at T=20mK.

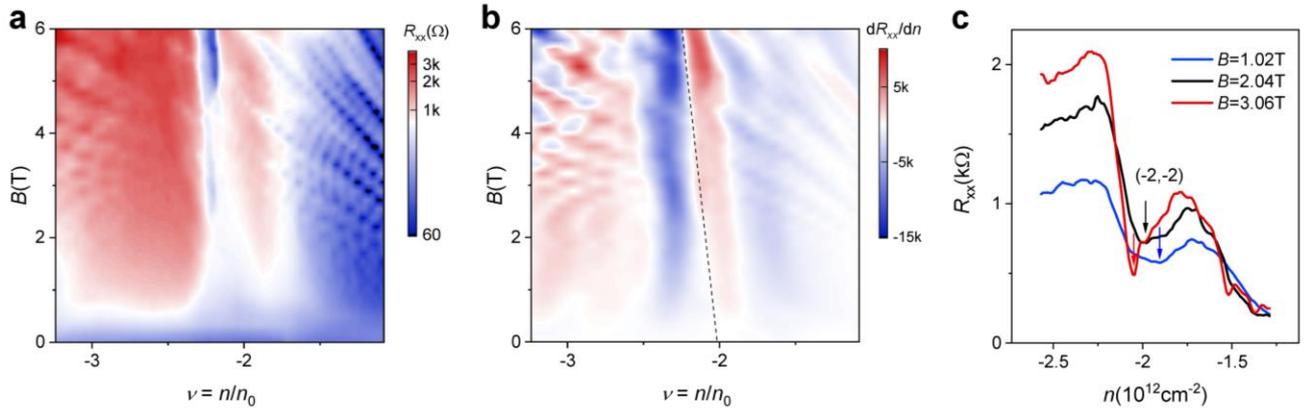

**FIG.S6** Onset of Chern insulator (-2, -2) in tiny magnetic fields for $\theta=1.25°$ TBG. a, zoomed-in Landau fan diagram. b, derivate of $R_{xx}$ with respect to carrier density n as a function of filling $v$ and field B. c, line cuts of $R_{xx}$ with respect to carrier density. All the data are acquired at T=20mK.

## 4. Landan fan diagram for varied twist angle

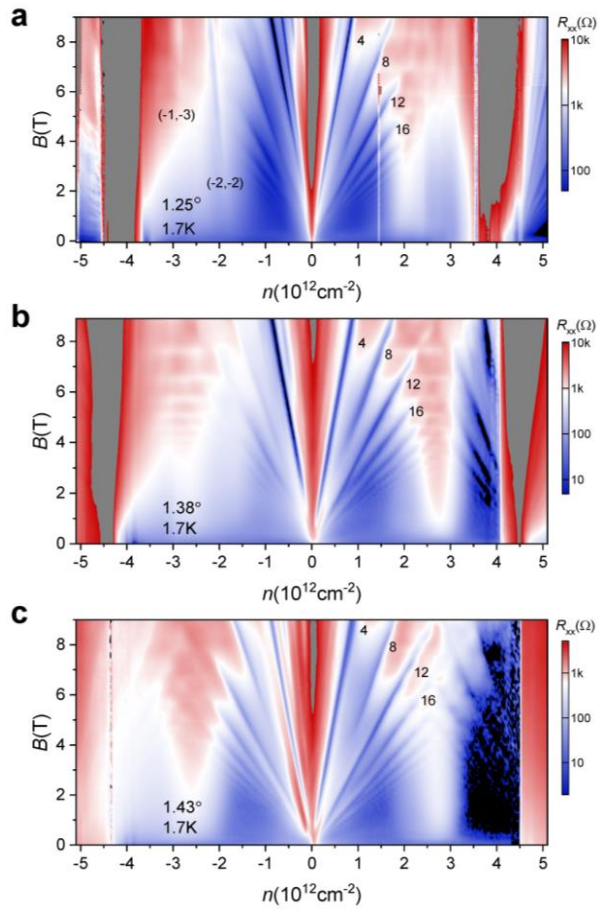

**FIG.S7** Landau fan diagram for both of hole and electron branches. The spike line in (a) comes from an unexpected error in our measurement system.

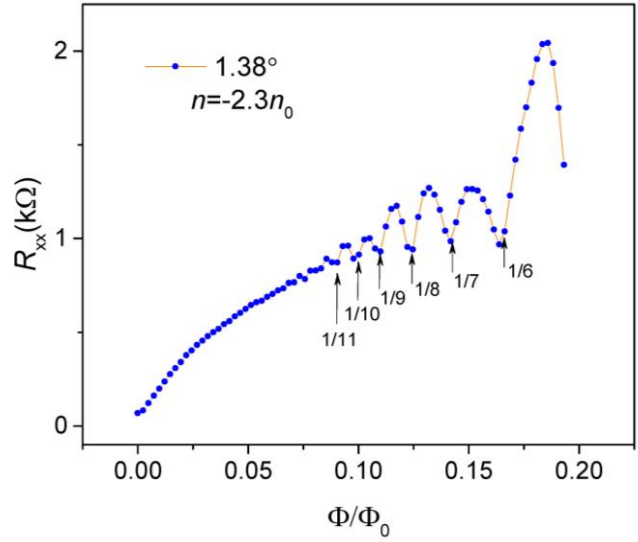

**FIG.S8** Brown-Zak oscillation in $\theta$=1.38° TBG. The data are obtained from Fig.S7b, shown as a line cut at density n=-2.3$n_0$.

We show a full image of Landau fan diagram both for electron and hole sides. At T=1.7K and B<9T, Chern insulators (-1, -3) and (-2, -2) still survive in the $\theta$=1.25° TBG. We observed more pronounced Brown-Zak oscillations at electron side as compared to one at T=20mK. This behavior is similar to that in graphene/hBN superlattice, where the fractal minibands will evolve into Brown-Zak subbands when temperature is elevated. Fig. S8 shows clearly Brown-Zak oscillations surviving down to $\phi_0$/11 in 1.38° TBG at hole side. While for 1.43° TBG, no signal of Brown-Zak oscillations is found.

## 5. Coulomb interaction and kinetic energy

A rigid calculation for Chern band formation at varied twist angle with e-e interaction involved is far beyond the content of our paper, instead we present a general analysis considering the competition between Coulomb interaction and kinetic energy here. In TBG with a small twist angle of θ, the Coulomb interaction can be estimated as $U = \frac{e^2}{4\pi\epsilon\epsilon_0\lambda}$, where e is the electron charge, $\epsilon$ the relative dielectric constant, $\epsilon_0$ the vacuum permittivity, and $\lambda \approx \frac{a}{\theta}$ the moire wavelength and $a$ the graphene lattice constant. We calculated the Coulomb interaction by assuming $\epsilon$=4 as shown in the picture below. From the continuum model, we extract bandwidth W for valence and conduction band to describe the kinetic energy. The bandwidth of valance band and conduction band is calculated by continuum model. We choose the minimum and maximum points on the path of mini-Brillouin zone from Gamma to M to K to calculate the bandwidth. Therefore, the bandwidth may slightly different from that calculated from the density of the state diagram. The calculation result indicates a crossover at $\theta = 1.27°$ when U/W=1, which locates the similar range of twist angle where we observed the symmetry-broken Chern insulators.

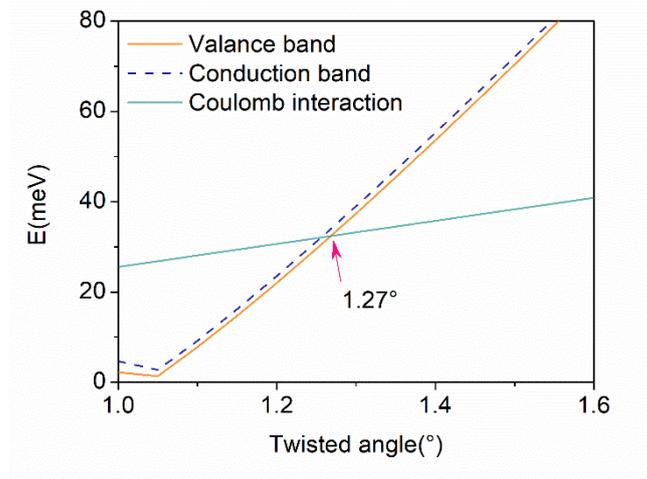

**FIG.S9** Band width calculated from continuum model and Coulomb repulsion energy for a small twist angle

It is also noted that the dielectric environments have a strong influence on the Coulomb energy. In our analysis, we chose a relative dielectric constant of 4, which is very close to that of hBN. The agreement between this simple assumption and our experimental results points to a screening effect in 3 dimensions by the dielectric environment. In experiments, $\epsilon$ might vary from device to device and bring a big uncertainty when quantifying Coulomb energy.